\documentstyle[epsfig]{aprim10}
\input{epsf}

\newif\ifAMStwofonts

\ifoldfss
  \ifCUPmtlplainloaded \else
    \NewTextAlphabet{textbfit} {cmbxti10} {}
    \NewTextAlphabet{textbfss} {cmssbx10} {}
    \NewMathAlphabet{mathbfit} {cmbxti10} {} 
    \NewMathAlphabet{mathbfss} {cmssbx10} {} 
  \fi
  \ifAMStwofonts
    \ifCUPmtlplainloaded \else
      \NewSymbolFont{upmath} {eurm10}
      \NewSymbolFont{AMSa} {msam10}
      \NewMathSymbol{\upi}     {0}{upmath}{19}
      \NewMathSymbol{\umu}     {0}{upmath}{16}
      \NewMathSymbol{\upartial}{0}{upmath}{40}
      \NewMathSymbol{\leqslant}{3}{AMSa}{36}
      \NewMathSymbol{\geqslant}{3}{AMSa}{3E}

       \let\le=\leqslant
       
    \fi
  \fi
\fi 

\ifnfssone
  \newmathalphabet{\mathit}
  \addtoversion{normal}{\mathit}{cmr}{m}{it}
  \addtoversion{bold}{\mathit}{cmr}{bx}{it}
  \newmathalphabet{\mathbfit} 
  \addtoversion{normal}{\mathbfit}{cmr}{bx}{it}
  \addtoversion{bold}{\mathbfit}{cmr}{bx}{it}
  \newmathalphabet{\mathbfss} 
  \addtoversion{normal}{\mathbfss}{cmss}{bx}{n}
  \addtoversion{bold}{\mathbfss}{cmss}{bx}{n}
  \ifAMStwofonts
    \ifCUPmtlplainloaded \else
      \UseAMStwoboldmath
      \makeatletter
      \new@mathgroup\upmath@group
      \define@mathgroup\mv@normal\upmath@group{eur}{m}{n}
      \define@mathgroup\mv@bold\upmath@group{eur}{b}{n}
      \edef\UPM{\hexnumber\upmath@group}
      \new@mathgroup\amsa@group
      \define@mathgroup\mv@normal\amsa@group{msa}{m}{n}
      \define@mathgroup\mv@bold\amsa@group{msa}{m}{n}
      \edef\AMSa{\hexnumber\amsa@group}
      \makeatother
      \mathchardef\upi="0\UPM19
      \mathchardef\umu="0\UPM16
      \mathchardef\upartial="0\UPM40
      \mathchardef\leqslant="3\AMSa36
      \mathchardef\geqslant="3\AMSa3E

       \let\le=\leqslant

    \fi
  \fi
\fi 

\ifnfsstwo
  \DeclareMathAlphabet{\mathbfit}{OT1}{cmr}{bx}{it}
  \SetMathAlphabet\mathbfit{bold}{OT1}{cmr}{bx}{it}
  \DeclareMathAlphabet{\mathbfss}{OT1}{cmss}{bx}{n}
  \SetMathAlphabet\mathbfss{bold}{OT1}{cmss}{bx}{n}
  \ifAMStwofonts
    \ifCUPmtlplainloaded \else
      \DeclareSymbolFont{UPM}{U}{eur}{m}{n}
      \SetSymbolFont{UPM}{bold}{U}{eur}{b}{n}
      \DeclareSymbolFont{AMSa}{U}{msa}{m}{n}
      \DeclareMathSymbol{\upi}{0}{UPM}{"19}
      \DeclareMathSymbol{\umu}{0}{UPM}{"16}
      \DeclareMathSymbol{\upartial}{0}{UPM}{"40}
      \DeclareMathSymbol{\leqslant}{3}{AMSa}{"36}
      \DeclareMathSymbol{\geqslant}{3}{AMSa}{"3E}

       \let\le=\leqslant

    \fi
  \fi
\fi 

\ifCUPmtlplainloaded \else
  \ifAMStwofonts \else 
    \def\upi{\pi}
    \def\umu{\mu}
    \def\upartial{\partial}
  \fi
\fi

\title[Galaxies and Environments]{Galaxies and Environment of the Clusters of Galaxies CL 0024+1654 and RX J0152.7-1357}

\author[Nugroho and Premadi]
       {Dading Nugroho$^1$ and Premana Premadi$^{1,}$$^2$\\
        $^1$Departement of Astronomy, Institut teknologi Bandung, Indonesia\\
        $^2$Bosscha Observatory, Lembang, Indonesia}
\date{}

\pagerange{\pageref{firstpage}--\pageref{lastpage}}
\pubyear{2008}

\begin{document}

\maketitle

\label{firstpage}

\begin{abstract}
We present the analysis and results of photometric and spectroscopic catalog combined with X-ray data of two non-relaxed clusters CL 0024+1654 (z=0.4) and RX J0152.7-1357 (z=0.8). Using the Spearman correlation analysis we quantify the correlation between morphology, color, and star formation rate of each galaxy with its surrounding number density, mass density, and temperature of Intracluster Medium (ICM). Although our results show that the two clusters exhibit a weaker correlation compared with relaxed clusters, it still confirms the significant effect of the ICM in varying the star formation rates in the galaxies. Various physical mechanisms have been suggested to explain the relation between the properties of galaxies and their environments for example: ram pressure stripping, mergers etc. Nonetheless, using this analysis alone, it is difficult to identify the dominant environmental mechanism(s) operating in clusters of galaxies and the role of the initial condition.
\end{abstract}


\section{Introduction}

A cluster of galaxies is in general defined as a self-gravitating system of galaxies that may be identified as a system exhibiting one or more of the following features: a higher concentration of galaxies relative to the average distribution of large scale structure of the universe;  massive concentration(s) of X-ray emitting gas, often associated with a   Sunyaev-Zel’dovich effect on the Cosmic Microwave Background Radiation; a non-random gravitational lensing shear map of background galaxies; and the largest gravitationally bound halo of dark matter in computer simulation of the large scale structure. A cluster provides a large number of galaxies at the same redshift within a relatively small projected field, suggesting that those galaxies might be gravitationally bound and might share a common history. The proximity of one galaxy to another, and also to the Intracluster Medium (ICM) suggests that environmental effect on galaxy evolution is plausible. In short, clusters of galaxies have become resourceful laboratories in studying the formation and evolution of galaxies, and thus the evolution of the large scale structure of the universe.

In general the components of a cluster are identified as galaxies, unbound stars recognised as the Intracluster Light, hot gas (Intracluster Medium, ICM), and dark matter. A supermassive blackhole may reside at the centre of a massive and relaxed cluster, giving rise to additional mechanisms to be considered in the cluster evolution. Each component may interact with one another, and there are suggested mechanisms for each interaction. The interactions are in general grouped into three, considering the participants: (1) interaction between ICM and galaxies, which involves the gaseous component of galaxies in direct contact with ICM, (2) interaction between the cluster potential and the galaxies, (3) interaction between galaxies.

All those interactions may affect the internal structure of the involved galaxies indicated by the significantly different values of some physical parameters of the galaxies (e.g. star formation rate) compared to those of isolated galaxies which experience zero or minimal environmental effect throughout their evolution.  The work presented in the conference concentrates on identifying the properties of each galaxy belonging to the cluster, identifying the environment of each of those galaxies, and studying any correlation between the properties of individual galaxies and their environments. The primary properties of a galaxy are taken to be its colour, morphology, and star formation rate. The environment is characterised by its surrounding galaxy number density, ICM temperature, and mass density.

\section{Methods}

The X-ray data were reduced and analyzed using XMM-Science Analysis Software (SAS). We applied the standard procedure in handling the XMM-Newton data, particularly the filtering process. The outcomes are ICM fluxes and temperature distribution. Smoothed X-ray images of two clusters are shown in Figures 1 and 12. 
\begin{figure}
\centering
\includegraphics[scale=0.6]{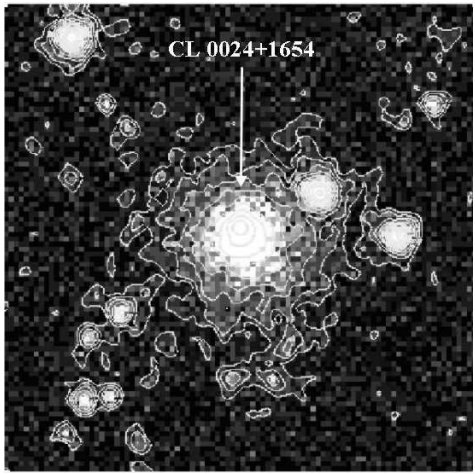}
\caption{X-ray image of CL 0024+1654 from XMM-Newton observations. North is up and east is to the left. The size of this field is 6.8$\prime$ $\times$ 6.8$\prime$ (2.08 Mpc $\times$ 2.08 Mpc at  z $\simeq$ 0.4)}\label{fig:cl0024}
\end{figure}

\begin{figure}
\centering
\includegraphics[scale=0.6]{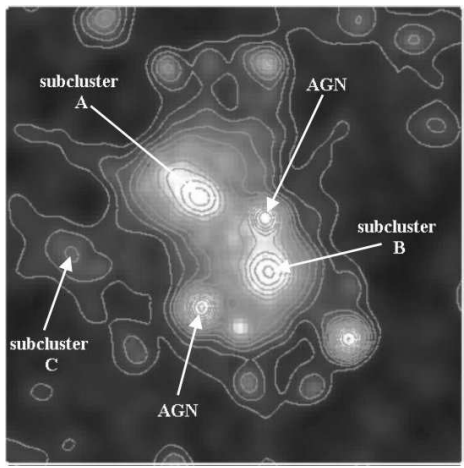}
\caption{X-ray image of RX J0152.7-1357. The field of view of this figure is 6.8$\prime$ $\times$ 6.8$\prime$ (3.02 Mpc $\times$ 3.02 Mpc at z $\simeq$ 0.8)}\label{fig:cl0152}
\end{figure}

We make use of information on the morphology of each galaxy given by Treu et al. \shortcite{treu03} for CL0024+1654, and by Blakeslee et al. \shortcite{blakeslee06} for RXJ 0152.7-1357. The morphology of each cluster galaxy is determined by applying two classification schemes, i.e. the Medium Deep Survey scheme for CL 0024 +1654 and the Hubble T type for RX J0152.7-1357. The Medium Deep Survey Scheme assigns morphology to accord with each class in Hubble Tuning Fork: -2=star, -1=compact, 0=ellipse,1=ellipse/lenticular(S0), 2=lenticular(S0), 3=Sa+b, 4=Spiral, 5=Sc+d, 6=irregular, 7=unclassified, 8=merger. With similar objectives, the Hubble T type has been used to classify the morphology of galaxies. \shortcite{postman05} visually inspected the galaxies in RX J0152.7-1357 to determine their morphologies and defined three broad categories: elliptical:-5$\le$T$\le$-3, lenticular: -2$\le$T$\le$0, and spiral+irregular: 1$\le$T$\le$10. 
 
We use optical color for both clusters to measure the color of the galaxies. We use (B-V) for CL 0024+1654 based on observation by the Canada France Hawaii Telescope \cite{czoske01} and (r-z) for RX J0152.7-1357 using Advanced Camera for Survey (ACS) aboard Hubble Space Telescope \cite{blakeslee06}.

As an indicator for star formation rate (SFR) we take O[II] emission line only, which may result in incompleteness in the SFR statistics. For further studies we should include the star formation indicators in ultraviolet and infrared bands as well.

In order to measure ICM temperature we analysed the spectra using XSPEC, an X-Ray Spectral Fitting Package. We fitted the filtered X-ray data with thermal plasma models using absorbed MEKAL model in energy range from 0.3 to 10 keV excluding 1.4-1.6 keV region due to Al profile and 7.45-9 keV region due to Cu line. MEKAL model is used to describe emission spectrum of hot diffuse gas based. There are five parameters in MEKAL model namely, plasma temperature in keV, hydrogen density, metal abundance, redshift. The fifth parameter is a normalization factor that one can choose to calculate or interpolate from model spectrum. We combined this model with absorption model following a local HI observation \cite{dickey90}. We also used redshift data for each of the cluster to fix the values of physical dimensions. The chemical the abundances are fixed to solar values. The output of the spectral fitting is temperature of the ICM. 

The environment of each galaxy in both clusters is studied, and a galaxy is termed “the target galaxy” when its environment is being considered. The number density, $\Sigma_{10}$, of each environment is calculated within a circular area of radius $R_{10}$, which is the distance to the farthest galaxy among the nearest 10 neighbors of the target galaxy. We also calculate the physical distance of each target galaxy from the center of the cluster. ICM, being the more diffuse environment, is characterised by its temperature and the flux of its X-ray emission. 

We then evaluated the relationship between color, morphology, and star formation rate of each galaxy with its environment using Spearman rank correlation analysis, which although limited to linear correlation, it is adequate for the purpose of examining general trends in the correlations.

\section{Results and Discussion}

The Spearman’s rank correlation coefficients show that the correlation between galaxies’ properties and their environments are weaker compared to those of more relaxed clusters, but the directions agree with expectation. We redid the correlation analysis separately for each type of galaxy and found that the strength of the correlations differ for the two types (shown in Figure 5). It indicates that the environmental mechanisms may operate differently for early and late type galaxies or operate at different stages of their evolutions.

\begin{figure}
\centering
\includegraphics[scale=0.50]{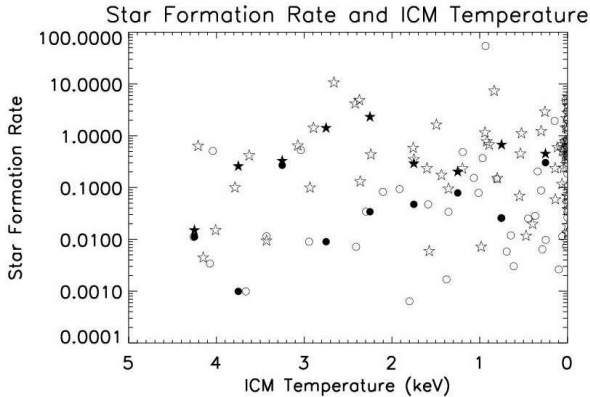}
\caption{Relation between star formation rate and ICM temperature in CL 0024+1654. Early type galaxies are denoted by stars and the late type by circle. Solid symbols indicate the median for each galaxy type.
}\label{fig:sfr_icm_0024}
\end{figure}

\begin{figure}
\centering
\includegraphics[scale=0.4]{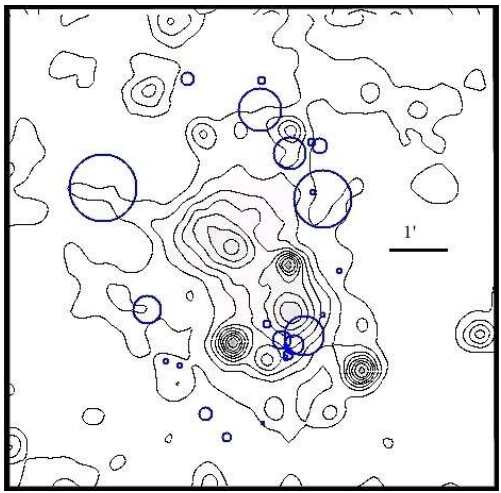}
\caption{Spatial distribution of star forming galaxies (indicated by circles) with respect to ICM distribution in RX J0152.7-1357. The size of the circle is proportional to the magnitude of the star formation rate. The 1$^{\prime}$ bar corresponds with 3.06 kpc at the cluster’s redshift.
}\label{fig:0152_sfr_position.jpeg}
\end{figure}

The most interesting result found in our analysis the correlation between star formation rate and ICM temperature (Figures 3 and 4) is that ICM pressure seemed to be able to enhance the star formation rate of the galaxies near the virial radius, but to decrease the star formation rate for galaxies towards the cluster core. This trend supported by detailed analysis of spectral classification of each galaxies which showed that near the virial radius there are large number of E+A (post-starburst) galaxies with strong Balmer absorption lines combine with absence of emission lines (e.g. [O II] or $H \alpha$). 

Environmental mechanisms that could give rise to such effects are ram pressure stripping and starvation by the ICM, and/or tidal compression by the cluster potential. The information is entangled that it is nearly impossible to single out a particular mechanism. Nonetheless, in this work we attempt to evaluate the role of only ICM in influencing star formation rate. This preliminary study calls for another series of examination on parameters directly related with the ICM-galaxy interaction. 

In the process of virialization, galaxies move about and inward in the cluster, and encounter various things along their courses. A galaxy moving towards and through a bulk of ICM would experience pressure exerted by the ICM. Depending on how the pressure is balanced by the galaxy potential, a portion of cool gas might be stripped of the galaxy. The incoming pressure depends on the velocity with which the galaxy encounters the ICM, the density of the ICM, and the cross section of the encounter. 

However, at this stage we could not yet determine the relative contributions from other mechanisms (merger and harassment) and the intrinsic property of galaxies which may provide the information of the initial condition. It is also imperative to recalculate the Star Formation Rate using complete indicators.

\section*{Acknowledgments}
We gratefully acknowledge the travel support from The Leids Kerkhoven-Bosscha Fonds (LKBF) and The International Astronomical Union (IAU). This research has made use of the NASA/IPAC Extragalactic Database (NED) which is operated by the Jet Propulsion Laboratory, California Institute of Technology, under contract with the National Aeronautics and Space Administration; the VizieR catalogue access tool, CDS, Strasbourg, France; NASA's Astrophysics Data System; and data from the High Energy Astrophysics Science Archive Research Center (HEASARC), provided by NASA's Goddard Space Flight Center.

\label{lastpage}

\clearpage

\end{document}